\documentstyle[prl,aps,mathrsfs,graphicx]{revtex}
\begin{document}
\draft

\def\overlay#1#2{\setbox0=\hbox{#1}\setbox1=\hbox to \wd0{\hss #2\hss}#1%
\hskip -2\wd0\copy1} \twocolumn[
\hsize\textwidth\columnwidth\hsize\csname@twocolumnfalse\endcsname

\title{Theory of a Slow-Light Catastrophe}
\author{Ulf Leonhardt}
\address{School of Physics and Astronomy, University of St Andrews,
North Haugh, St Andrews, Fife, KY16 9SS, Scotland}

\maketitle
\begin{abstract}
In diffraction catastrophes such as the rainbow the wave nature
of light resolves ray singularities and draws delicate
interference patterns. In quantum catastrophes such as the black
hole the quantum nature of light resolves wave singularities and
creates characteristic quantum effects related to Hawking
radiation. The paper describes the theory behind a recent
proposal {[}U. Leonhardt, arXiv:physics/0111058, Nature (in
press){]} to generate a quantum catastrophe of slow light.
\end{abstract}
\date{\today}

\vskip2pc]

\narrowtext

\section{Introduction}

Catastrophes \cite{BerryUpstill} are at the heart of many
fascinating optical phenomena. The most prominent example of such
a catastrophe is the rainbow. Light rays from the Sun enter water
droplets floating in the air. After two refractions and one
reflection inside each drop the rays reach an observer. Above a
critical observation angle no rays arrive, whereas below the
angle two rays strike the observer. A bright bow, the rainbow,
appears at the critical angle, because here the cross section of
light rays diverges \cite{Nussenzveig}. (The critical angle
depends on the refractive index  that varies with the frequency
of light in dispersive media such as water, giving rise to the
rainbow colors.) The direction of a light ray is proportional to
the gradient of the phase. The rainbow thus represents a
singularity of a gradient map, a catastrophe in the sense of Thom
\cite{Thom} and Arnol'd \cite{Arnold}. Structurally stable
singularities of gradient maps fall into distinct classes,
depending on the number of control parameters involved
\cite{Thom,Arnold}. Structural stability is the key to Nature's
way of focusing light \cite{Nye} in the caustics created by ray
catastrophes. Yet the wave nature of light smooths the harsh
singularities of rays. Simultaneously, characteristic
interference effects appear. For example, the pairs of light rays
below the rainbow create a delicate pattern of supernumerary arcs
\cite{BerryUpstill} that are visible under favorable weather
conditions (when the floating droplets are nearly uniform in size
\cite{Nussenzveig}). Every class of diffraction catastrophes
generates its distinct interference structure \cite{BerryUpstill}.

Catastrophe optics describes the wave properties of ray
singularities. In the hierarchy of physical concepts, wave optics
refines and embraces ray optics, and quantum optics rules above
wave optics. So, what would be the quantum effects of wave
catastrophes \cite{Berry}? First, what are quantum catastrophes?
It might be a good idea to begin with an example, the black hole
\cite{Misner}. When a star collapses to a black hole an event
horizon is formed, cutting space into two disconnected regions.
Seen from an outside observer, time stands still at the horizon,
freezing all motion. A light wave would freeze as well,
propagating with ever-shrinking wavelength. In mathematical terms
\cite{Brout}, a monochromatic light wave of frequency $\omega$
oscillates as $\Theta(r-r_s)(r-r_s)^{i\mu}$ when the radius $r$
approaches the horizon $r_s$, with $\mu=2r_s\,\omega/c$. A
logarithmic phase singularity will develop. Potential quantum
effects of such a wave singularity are effects of the quantum
vacuum. The gravitational collapse \cite{Misner} of the star into
the black hole has swept along the vacuum. The vacuum thus shares
the fate of an inward-falling observer. Yet such an observer
would not notice anything unusual at the event horizon. In
mathematical terms, the vacuum modes are analytic across the
horizon \cite{Brout,Unruh}. On the other hand, the modes
perceived by an outside observer are essentially non-analytic,
because they vanish beyond the horizon where the observer has no
access to. Consequently, the observer does not see the
electromagnetic field in the vacuum state. Instead, the observer
notices the quanta of Hawking radiation \cite{Hawking} with the
Planck spectrum $(e^{2\pi\mu}-1)^{-1}$. The quantum vacuum does
not assume catastrophic waves of the type
$\Theta(r-r_s)(r-r_s)^{i\mu}$, hence resolving so the associated
wave singularity and, simultaneously, generating quantum radiation
with a characteristic spectrum. At the heart of such a
catastrophe lies a time-dependent process, for example the
gravitational collapse in the case of the black hole
\cite{Misner}. The process has disconnected the spatial regions
where waves can propagate and has created a logarithmic phase
singularity at the interface. Any time-dependent phenomenon will
generate some radiation, as long as the process lasts. In
remarkable contrast, a quantum catastrophe creates quanta
continuously.

This paper describes the theory behind a recent idea
\cite{LeoNature} to generate a quantum catastrophe of slow light
\cite{Slow,Liu,Philips,Dutton,EIT,Harris,Fleisch,Moseley,Inouye}.
An experiment is proposed based on Electromagnetically-Induced
Transparency (EIT) \cite{EIT}. In EIT a control beam determines
the optical properties of slow light in a suitable medium.
Changing the intensity of the control light from a uniform to a
parabolic profile creates s slow-light catastrophe
\cite{LeoNature}, see Fig.\ 1. This catastrophe resembles the
event horizon of a black hole but also shows some characteristic
differences. In Section II we put forward a rather general
phenomenological quantum field theory of slow light. Appendix A
justifies the theory for the most common method to slow down
light using EIT \cite{EIT}. In Section III we address the
specific theory of the slow-light catastrophe \cite{LeoNature}.
Appendix B contains some estimations that are relevant to the
experimental aspects involved. Section IV summarizes the results
and draws a further vision of quantum catastrophes.

\begin{figure}[htbp]
\begin{center}
\includegraphics[width=20.5pc]{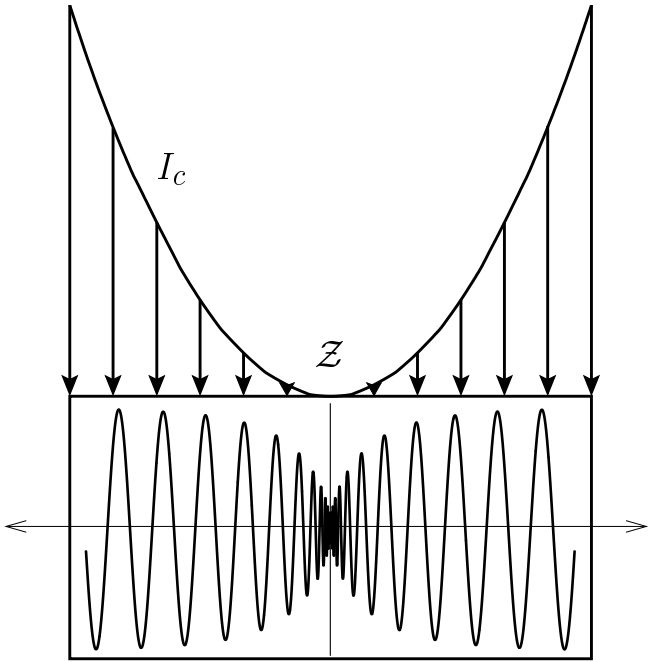}
\vspace*{2mm} \caption{Schematic diagram of the proposed
experiment. A beam of control light with intensity $I_c$ generates
Electromagnetically-Induced Transparency [15] in a medium,
strongly modifying its optical properties for a second field of
slow light. When an initially uniform control intensity is turned
into the parabolic profile shown in the figure, the slow-light
field suffers a quantum catastrophe. To slow-light waves, the
interface ${\cal Z}$ of zero control intensity cuts space into two
disconnected regions and creates a logarithmic phase singularity,
in analogy to the effect [7] of an event horizon [6]. The quantum
vacuum of slow light cannot assume such catastrophic waves. In
turn, pairs of slow-light quanta, propagating in opposite
directions away from ${\cal Z}$, are emitted with a
characteristic spectrum. The waves shown below the intensity
profile refer to the emitted light with the modes $w_R$ and $w_L$
of Eq.\ (\ref{set}). } \label{figure1}
\end{center}
\end{figure}

\section{The model}
Electromagnetically-Induced Transparency (EIT) \cite{EIT} has
served as a method to slow down light significantly \cite{Slow}
or, ultimately, to freeze light completely
\cite{Liu,Philips,Dutton,Turukhin}. Like other successful
techniques, EIT is based on a simple idea \cite{EIT}: A control
beam of laser light couples the upper levels of an atom, and, in
this way, the beam strongly modifies the optical properties of
the atom. In particular, the coupling of the excited states
affects the transition from the atomic ground state to one of the
upper states, {\it i.e.}, the ability of the atom to absorb probe
photons with matching transition frequency. Destructive quantum
interference between the paths of the transition process turns
out to eliminate absorption at exact resonance \cite{EIT}. A
medium composed of such optically-manipulated atoms is
transparent at a spectral line where it would otherwise be
completely opaque. In the vicinity of the transparency frequency
$\omega_0$ the medium is highly dispersive, {\it i.e.} the
refractive index changes within a narrow frequency interval. In
turn, probe light pulses with a carrier frequency at $\omega_0$
travel with a very low group velocity $v_g$ \cite{Harris}. The
intensity $I_c$ of the control beam determines the group velocity
of the probe pulse, as long as $I_c$ is stronger than the probe.
Paradoxically, the lower $I_c$ is the slower moves the pulse,
which, however, is only possible when the electronic states of the
atoms follow dynamically the control field \cite{Fleisch}, causing
the probe-light intensity to fall accordingly \cite{Fleisch}. In
this regime light freezes when $I_c$ vanishes
\cite{Liu,Philips,Dutton,Turukhin}.

The theory of EIT \cite{EIT} often employs an atomic three-level
scheme: Two levels account for the excited states coupled by the
control field and one level represents the ground state, see
Appendix A. In reality, atoms are more complicated and, when
details matter, an accurate description involves the full atomic
sub-level structure \cite{Moseley}. Here we put forward a
phenomenological quantum theory of slow light that is rather
independent of the microscopic mechanisms used in practice. We
assume only that the slow-light medium is transparent with a real
susceptibility that depends linearly on the detuning from
$\omega_0$. In EIT our model is restricted to the narrow
transparency window around $\omega_0$ that can maximally reach
the natural line width of the atomic transition. Appendix A shows
that our model agrees with the three-level theory of EIT. Our
theory is simple enough to be treated analytically and yet
sufficiently complex to capture the essence of slow-light quanta.
We postulate an effective Lagrangian ${\mathscr L}$, show that
${\mathscr L}$ is consistent with the known dynamics of EIT within
the validity range of our model, calculate the energy and
quantize the field of slow-light polaritons. Our approach has the
additional advantage that it may be applicable to other
mechanisms \cite{Inouye} for creating slow light that do not rely
on EIT.

\subsection{Lagrangian}

We characterize slow light by a real scalar field $\varphi$
ignoring the polarization. The optical field $\varphi$ shall be
given in units of the vacuum noise \cite{MandelWolf}. For
simplicity, we assume uniformity in two spatial directions, $x$
and $y$, and regard the optical field as a function of time $t$
and position $z$. Throughout this paper we denote partial time
derivatives by dots, by differential operators $\partial_t$ or
simply by suffixes $t$, whatever happens to be most convenient.
Spatial derivatives are denoted by dashes, differential operators
$\partial_z$ or suffixes $z$. We postulate the effective
Lagrangian
\begin{equation}
\label{lagrangian} {\mathscr L} = \frac{\hbar}{2}\left((1 +
\alpha)\dot{\varphi}^2 - c^2\varphi'^2 -
\alpha\,\omega_0^2\varphi^2\right) \,.
\end{equation}
We see in the next subsection that the real parameter function
$\alpha$ determines the group velocity, and hence we call $\alpha$
group index. In EIT \cite{EIT} the parameter $\alpha$ is inversely
proportional to the control-field intensity,
\begin{equation}
\label{control} \alpha(t,z) = \frac{\kappa}{I_c(t,z)} \,,
\end{equation}
with a coupling strength $\kappa$ that is proportional to the
modulus squared of the atomic dipole-transition matrix element and
to the number of atoms per unit volume. Without the EIT medium
present, ${\mathscr L}$ is the Lagrangian of a free
electromagnetic field, $\frac{\varepsilon_0}{2}(E^2-c^2B^2)$ with
fixed polarization. We see that $\varphi$ is related to the
electric field strength $E$ in SI units by
\begin{equation}
E = \left(\frac{\hbar}{\varepsilon_0}\right)^{1/2}\omega\, \varphi
\end{equation}
where $\omega$ denotes the frequency of light and $\varepsilon_0$
is the electric permittivity of the vacuum.

In order to motivate the Lagrangian (\ref{lagrangian}) we
consider the corresponding Euler-Lagrange equation
\cite{Weinberg},
\begin{equation}
\partial_t \frac{\delta {\mathscr L}}{\delta\dot{\varphi}} +
\partial_z \frac{\delta {\mathscr L}}{\delta\varphi'} =
\frac{\delta {\mathscr L}}{\delta\varphi} \,,
\end{equation}
that leads to the wave equation
\begin{equation}
\label{wave} \left(\partial_t(1+\alpha)\partial_t - c^2
\partial_z^2 + \alpha\,\omega_0^2\right) \varphi = 0 \,.
\end{equation}
This is the propagation equation of slow light based on the
traditional three-level model, see Appendix A, but the equation
holds on more general grounds: Equation (\ref{wave}) describes
light in media with linear spectral susceptibility. Assuming that
the optical field oscillates at much shorter time and length
scales than any variations of $\alpha$, we replace $i\partial_t$
by the frequency $\omega$ and $-i\partial_z$ by the wave number
$k$. We arrive at the dispersion relation
\begin{equation}
\label{dispersion} k^2 - \frac{\omega^2}{c^2} - \alpha\,
\frac{(\omega+\omega_0)(\omega-\omega_0)}{c^2} = 0
\end{equation}
which, in the positive transparency window near $\omega_0$, agrees
with
\begin{equation}
k^2 - \frac{\omega^2}{c^2}(1+\chi) = 0
\end{equation}
and the linear spectral susceptibility \cite{Harris,LPliten}
\begin{equation}
\label{chi1} \chi = \frac{2\alpha}{\omega_0}\,(\omega-\omega_0)
\,.
\end{equation}
At a later stage we need to consider negative frequencies as well.
To verify that the dispersion relation (\ref{dispersion}) is also
valid in the corresponding negative transparency window we
utilize the general property of a spectral susceptibility
\begin{equation}
\chi(-\omega) = \chi^*(\omega)
\end{equation}
which implies near $-\omega_0$
\begin{equation}
\label{chi2} \chi = -\frac{2\alpha}{\omega_0}\,(\omega+\omega_0)
\,.
\end{equation}
We see that as long as the frequency of the probe light lies
within the transparency windows of EIT, the Lagrangian
(\ref{lagrangian}) reproduces the typical linear slope of the
spectral susceptibility. In fact, up to a trivial prefactor,
${\mathscr L}$ is the only Lagrangian that is quadratic in the
field and its derivatives and that is consistent with the
spectral susceptibilities (\ref{chi1}) and (\ref{chi2}). Therefore
we regard ${\mathscr L}$ as a suitable Lagrangian for slow light.

\subsection{Dynamics}

The ability to freeze light by turning off the control field
depends crucially on the dynamics of the process. Consider a
time-dependent group index $\alpha$ without significant spatial
variations. In this case slow light is dominated by oscillations
within an optical wave length and an optical cycle. We express
the wave as
\begin{equation}
\label{ansatz}
\left(\frac{\hbar}{\varepsilon_0}\right)^{1/2}\omega\, \varphi =
{\cal E} \, e^{\displaystyle ikz-i\omega t} + {\rm c.c.}\,,\quad
k = \frac{\omega}{c}\,,
\end{equation}
with the slowly varying electric-field amplitude ${\cal E}$ in SI
units. We approximate
\begin{eqnarray}
\left(\frac{\hbar}{\varepsilon_0}\right)^{1/2}\omega\,
e^{\displaystyle -ikz+i\omega t}\,\, \ddot{\varphi} &\approx&
\left(-\omega^2 -
2i\omega\partial_t \right) {\cal E}\,,\nonumber\\
\left(\frac{\hbar}{\varepsilon_0}\right)^{1/2}\omega\,
e^{\displaystyle -ikz+i\omega t}\,\, \dot{\varphi} &\approx&
-i\omega {\cal E} \,,
\nonumber\\
\left(\frac{\hbar}{\varepsilon_0}\right)^{1/2}\omega\,
e^{\displaystyle -ikz+i\omega t} \varphi'' &\approx& \left(-k^2 +
2ik\partial_t \right) {\cal E}\,,
\end{eqnarray}
and get from the wave equation (\ref{wave})
\begin{eqnarray}
\lefteqn{-2i\omega(1+\alpha)\dot{{\cal E}}} \nonumber\\ & &\approx
\left((1 + \alpha)\omega^2 + i\omega\dot{\alpha} - c^2k^2 +
2ikc^2\partial_z - \alpha\omega_0^2 \right) {\cal E} \nonumber\\
& & = \left(2i\omega c \partial_z + i\omega\dot{\alpha} +
\alpha(\omega^2 - \omega_0^2) \right) {\cal E} \nonumber\\
& & = 2i\omega\left(c{\cal E}' + \frac{\dot{\alpha}}{2}{\cal E}
\right)
\end{eqnarray}
when the carrier frequency $\omega$ is equal to the transparency
resonance $\omega_0$. We apply the relation (\ref{control})
between the control-field intensity $I_c$ and the group index
$\alpha$, write $I_c$ as the square of the field strength ${\cal
E}_c$, and obtain, finally,
\begin{equation}
\label{ee} \dot{{\cal E}} + c{\cal E}' = -\alpha\dot{{\cal E}} -
\frac{\dot{\alpha}}{2} {\cal E} = - \frac{\kappa}{{\cal
E}_c}\,\partial_t \frac{{\cal E}}{{\cal E}_c} \,.
\end{equation}
This is exactly the propagation equation of slow light in the
adiabatic and perturbative limit (Eq.\ (9) of Ref.\
\cite{Fleisch} with the Rabi frequency $\Omega$ being
proportional to ${\cal E}_c$). Consequently, the Lagrangian
(\ref{lagrangian}) has codified naturally the correct dynamic
regime including the $\frac{1}{2}\dot{\alpha}{\cal E}$ term that
describes reversible stimulated Raman scattering \cite{Fleisch}.

In order to understand the principal behavior of ordinary
slow-light pulses, consider the case of a spatially uniform yet
time-dependent group index. Equation (\ref{ee}) has the solution
\begin{equation}
\label{intensity} {\cal E}(t,z) = {\cal E}_0\Big(z - \int\! v_g\,
dt\Big)\,\sqrt{v_g/c}
\end{equation}
in terms of the velocity \cite{LPliten}
\begin{equation}
\label{group} v_g = \frac{c}{1 + \alpha} \,.
\end{equation}
We see that the pulse envelope ${\cal E}$ propagates with the
speed $v_g$ called group velocity. When the group velocity changes
in time, the intensity $|{\cal E}|^2$ reacts proportionally. The
ratio between the control (\ref{control}) and the pulse intensity
(\ref{intensity}), $(I_c + \kappa)/\,|{\cal E}_0|^2$, remains
large, even in the limit of a vanishing control field when $v_g$
vanishes as well, as long as $\kappa$ is large (for a
sufficiently dense medium). The spectral spread of a pulse is
reduced by $v_g/c$ and a standing pulse oscillates just with the
carrier frequency. In this regime slow light does not leave the
transparency window of EIT \cite{Philips}. One can freeze light
without losing control \cite{Liu,Philips}.

\subsection{Energy}

After having gained confidence in our field-theoretical approach,
we use the Lagrangian (\ref{lagrangian}) to calculate the energy
balance of slow light. According to Noether's theorem
\cite{Weinberg} we obtain the energy density
\begin{equation}
I = \frac{\delta {\mathscr L}}{\delta
\dot{\varphi}}\,\dot{\varphi} - {\mathscr L} =
\frac{\hbar}{2}\left((1 + \alpha)\dot{\varphi}^2 + c^2\varphi'^2 +
\alpha\,\omega_0^2\varphi^2\right)
\end{equation}
and the energy flux (Poynting vector)
\begin{equation}
P = \frac{\delta {\mathscr L}}{\delta \varphi'}\,\dot{\varphi} =
-\hbar c^2 \dot{\varphi}\varphi' \,. \label{energy}
\end{equation}
The energy balance $I_t + P_z$ is then, as a consequence of the
wave equation (\ref{wave}),
\begin{equation}
I_t + P_z = \frac{\hbar\dot{\alpha}}{2}\left(\dot{\varphi}^2 +
\omega_0^2\varphi^2\right) \,.
\end{equation}
Temporal changes in the control field, modifying the group index
(\ref{control}), do not conserve energy. In fact, the experiment
\cite{Liu} indicates that the control beam can amplify light
stored in an EIT medium with zero group velocity. In the
experiment \cite{Liu}, slow light enters the EIT sample and is
then frozen inside by turning off the control field. Switching on
the control releases the stored light. The pulse emerges with an
intensity that depends on the control field and that may exceed
the initial intensity, in agreement with Eq.\ (\ref{intensity}).
Clearly, this phenomenon is only possible if energy is indeed
transferred from the control beam to the probe light.

\subsection{Polaritons}

Finally, we realize the full potential of the Lagrangian
(\ref{lagrangian}) in setting up an effective quantum theory of
slow light. The classical canonical momentum density of the field
$\varphi$ is \cite{Weinberg,Birrell}
\begin{equation}
\frac{\delta{\mathscr L}}{\delta\dot{\varphi}} = \hbar(1 + \alpha)
\dot{\varphi} \,.
\end{equation}
We quantize the field by regarding $\varphi$ and $\delta{\mathscr
L}/\delta\dot{\varphi}$ as Hermitian operators $\hat{\varphi}$ and
$\hat{\pi}$, respectively, with the canonical commutation
relations \cite{Weinberg,Birrell}
\begin{eqnarray}
[\hat{\varphi}(t,z),\hat{\varphi}(t,z')] & = &
[\hat{\pi}(t,z),\hat{\pi}(t,z')] = 0 \,,\nonumber\\
\mbox{$[$}\hat{\varphi}(t,z),\hat{\pi}(t,z')\mbox{$]$} &=&
i\hbar\,\delta(z-z') \,.
\end{eqnarray}
Let us decompose the field $\hat{\varphi}$ into modes with
dimensionless mode indices $q$
\begin{equation}
\label{modes}\hat{\varphi}(t,z) = \int \left(\hat{a}_q u_q(t,z) +
\hat{a}_q^\dagger u_q^*(t,z)\right) dq\,.
\end{equation}
In order to guarantee that $\hat{\varphi}$ satisfies the wave
equation (\ref{wave}) the mode functions $u_q$ are required to
obey Eq.\ (\ref{wave}) as well. The $u_q$ shall be normalized
according to
\begin{equation}
\label{norm} \big(u_q,u_{q'}\big) = \delta(q-q')\,,\quad
\big(u_q,u_{q'}^*\big) = 0 \,,
\end{equation}
with the Klein-Gordon-type scalar product \cite{Weinberg,Birrell}
\begin{equation}
\label{scalar} \left(\varphi_1,\varphi_2\right) = i
\int_{-\infty}^{+\infty}\Big(\varphi_1^*\dot{\varphi}_2 -
\dot{\varphi}_1^*\varphi_2\Big)(1 + \alpha)\, dz \,.
\end{equation}
The scalar product is chosen such that it remains constant during
the propagation of $\varphi_1$ and $\varphi_2$,
\begin{eqnarray}
\partial_t \left(\varphi_1,\varphi_2\right) & = & i
\int_{-\infty}^{+\infty}\Big(\varphi_1^*\partial_t (1 + \alpha)
\dot{\varphi}_2 - \varphi_2\partial_t (1 +
\alpha)\dot{\varphi}_1^*\Big)\, dz \nonumber\\
& = & ic^2 \int_{-\infty}^{+\infty} \partial_z
\left(\varphi_1^*\varphi'_2 - \varphi_2{\varphi_1^*}'\right) dz
\nonumber\\
& = & 0 \,.
\end{eqnarray}
Using these postulates and definitions we calculate the
commutation relation of the mode operators
\begin{eqnarray}
[\hat{a}_q,\hat{a}_{q'}^\dagger] & = & -
\left(u_q,\hat{\varphi}\right)\left(u_{q'}^*,\hat{\varphi}\right)
+
\left(u_{q'}^*,\hat{\varphi}\right)\left(u_q,\hat{\varphi}\right)
\nonumber\\
& = & \int_{-\infty}^{+\infty}\left(u_q^*\hat{\varphi}_t -
\hat{\varphi}\dot{u}_q^*\right)(1 + \alpha)\, dz \nonumber\\
& & \times \int_{-\infty}^{+\infty}\left(u_{q'}\hat{\varphi}_t -
\hat{\varphi}\dot{u}_{q'}\right)(1 + \alpha)\, dz' \nonumber\\
& & - \int_{-\infty}^{+\infty}\left(u_{q'}\hat{\varphi}_t -
\hat{\varphi}\dot{u}_{q'}\right)(1 + \alpha)\, dz' \nonumber\\
& & \times \int_{-\infty}^{+\infty}\left(u_q^*\hat{\varphi}_t -
\hat{\varphi}\dot{u}_q^*\right)(1 + \alpha)\, dz \nonumber\\
& = &
\frac{1}{\hbar}\int_{-\infty}^{+\infty}\int_{-\infty}^{+\infty}
\Big(u_q^*\dot{u}_{q'}[\hat{\varphi}(z'),\hat{\pi}(z)]\left(1 +
\alpha(z')\right)\nonumber\\
& & -\quad\,
\dot{u}_q^*u_{q'}[\hat{\varphi}(z),\hat{\pi}(z')]\left(1 +
\alpha(z)\right)\Big)\, dz\,dz' \nonumber\\
& = & i\int_{-\infty}^{+\infty}\left(u_q^*\dot{u}_{q'} -
\dot{u}_q^*u_{q'}\right) (1 + \alpha)\, dz\nonumber\\
& = & \delta(q - q') \,. \label{comm}
\end{eqnarray}
In a similar way we prove that
\begin{equation}
[\hat{a}_q,\hat{a}_{q'}] = 0\,.
\end{equation}
Consequently, and in agreement with the spin-statistics theorem
\cite{Weinberg}, slow light consists of bosons. Let us express
the total energy (\ref{energy}) in terms of the annihilation and
creation operators $\hat{a}_q$ and $\hat{a}_q^\dagger$. Consider
the case of a stationary group index $\alpha$ when the total
energy is conserved. We obtain after partial integration, via the
wave equation (\ref{wave}) for the field operators,
\begin{eqnarray}
\lefteqn{\int_{-\infty}^{+\infty}\hat{T}^{00}\,dz} \nonumber\\
& & = \frac{\hbar}{2} \int_{-\infty}^{+\infty} \left((1 +
\alpha)(\partial_t \hat{\varphi})^2 -
c^2\hat{\varphi}\,\partial_z^2\hat{\varphi} +
\alpha\,\omega_0^2\hat{\varphi}^2\right) dz
\nonumber\\
& & = \frac{\hbar}{2} \int_{-\infty}^{+\infty} \left((\partial_t
\hat{\varphi})^2 - \hat{\varphi}\,\partial_t^2\hat{\varphi}
\right)(1 + \alpha)\,dz \,.
\end{eqnarray}
We employ the mode expansion (\ref{modes}) with the norm
(\ref{norm}) with respect to the scalar product (\ref{scalar}),
use the commutation relation (\ref{comm}), and find, finally,
\begin{equation}
\int_{-\infty}^{+\infty}\hat{T}^{00}\,dz = \int
\hbar\omega\left(\hat{a}_q^\dagger\hat{a}_q +
\frac{1}{2}\right)dq \,.
\end{equation}
Consequently, the annihilation and creation operators $\hat{a}_q$
and $\hat{a}_q^\dagger$ refer indeed to the energy quanta of slow
light. The quasiparticles of light in a dielectric medium are
called polaritons in analogy to the optical excitations in a solid
\cite{Kittel}. They combine characteristic features of free
photons with the properties of the dipole oscillations of the
atoms constituting the medium. When light is slowed down photons
turn into atomic excitations that, after acceleration, may emerge
as photons again \cite{Liu,Philips}. The polariton picture
contains implicitly the correct book-keeping of the photonic and
atomic features of light in linear media.

One can use similar arguments as in the subsection on the
dynamics of slow light to prove that the proposed polariton theory
is consistent with the adiabatic three-level model \cite{Fleisch}.
Yet our approach is not restricted to a regime dominated by
spatial oscillations of the form $\exp(i\omega z/c)$, see Eq.\
(\ref{ansatz}), which is in essence the regime of geometrical
optics \cite{BornWolf}. One can easily relax this unnecessary
constraint of the three-level theory, because the electromagnetic
response of an atom is local, as long as the wave length of light
is large compared with the size of the atom. The adiabatic
polariton theory \cite{Fleisch} is, as ours, restricted to a
narrow frequency range with respect to time, {\it i.e.}, to the
transparency window of the electromagnetically-manipulated
medium. Yet this restriction in frequency does not exclude rapid
spatial oscillations beyond the scale of the wave length in
vacuum, $2\pi c/\omega$. We will see in the next section that
such oscillations occur near a spatial singularity of the group
index. In this situation our quantum field theory of slow-light
polaritons turns into a perfect tool for analyzing the quantum
physics of a wave catastrophe.

\section{The catastrophe}

Imagine that the control beam illuminates the EIT medium from
above. Initially, the control intensity is uniform, but then the
control light develops a dark node that continues as an interface
${\cal Z}$ of zero intensity $I_c$ through a part of the medium.
One could use computer-generated holograms to achieve this
situation, similar to the generation of Laguerre-Gaussian beams
\cite{Padgett}. Suppose that the interface is sufficiently flat
and cuts deep enough into the medium to justify our model. (We
have assumed uniformity in the two spatial directions $x$ and $y$
that are parallel to the interface. Most probably, uniformity
over a few wave lengths would suffice.) Near a zero in the
intensity, the control field strength must grow linearly.
Consequently, $I_c$ depends quadratically on $z$ and, according
to the relation (\ref{control}), the group index $\alpha$ forms a
quadratic singularity where the group velocity (\ref{group})
vanishes,
\begin{equation}
\label{catastrophe} \alpha = \frac{a^2}{z^2}\,.
\end{equation}
The parameter $a$ sets the scale of the group-index profile. We
assume that the spatial profile (\ref{catastrophe}) of the group
index extends over a sufficiently long range. For simplicity, we
consider a one-dimensional model where the slow light propagates
in $z$ direction only. Appendix B generalizes our results to a
realistic three-dimensional situation. Apart from these
idealizations and from the physics captured in our Lagrangian
(\ref{lagrangian}), we make no further approximations.

\subsection{Horizon}

At a node of the control field the group velocity of the probe
light vanishes. Therefore we would expect that no wave packet of
slow light can pass the interface ${\cal Z}$. Consider slow light
subject to the wave equation (\ref{wave}) with the group-index
profile (\ref{catastrophe}). We represent a wave $\varphi(t,z)$ as
\begin{equation}
\varphi = \sqrt{z}\,\phi
\end{equation}
and find
\begin{equation}
\label{phiwave} \left[ z^2\partial_t^2 + a^2(\partial_t^2 +
\omega_0^2) - c^2\left(z\partial_z\,z\partial_z - {\textstyle
\frac{1}{4}}\right) \right] \phi = 0 \,.
\end{equation}
We could multiply $\phi$ by step functions $\Theta(\pm z)$ and
still get solutions of the wave equation (\ref{phiwave}), because
\begin{eqnarray}
z\partial_z\,\Theta(\pm z)\phi(z) & = & \Theta(\pm z)\,
z\partial_z\,\phi(z) \pm \phi(z)\,z\,\delta(z) \nonumber\\
& = &\Theta(\pm z)\,z\partial_z\,\phi(z) \,.
\end{eqnarray}
Consequently, waves on different sides of ${\cal Z}$ are
independent from each other. As long as slow light is concerned,
the interface ${\cal Z}$ has cut space into two disconnected
parts.

Consider monochromatic probe light oscillating with frequency
$\omega$. In this case the wave equation (\ref{phiwave}) reduces
to Bessel's differential equation \cite{Erdelyi2} with the index
\begin{equation}
\nu = \sqrt{{\textstyle\frac{1}{4}} - a^2(k^2-k_0^2)} \,.
\end{equation}
Here $k$ abbreviates $\omega/c$ and $k_0$ refers to $\omega_0/c$.
So, in mathematical terms, the monochromatic waves of slow light
are products of a square root with Bessel functions
\cite{Erdelyi2},
\begin{equation}
\label{phi} \varphi = \sqrt{z}\,J_{\pm\nu}(kz)\,e^{\displaystyle
-i\omega t} \,.
\end{equation}
Depending on the detuning of the frequency $\omega$ with respect
to the exact transparency resonance $\omega_0$, two cases emerge.
First, when $4a^2(k^2-k_0^2)$ is below unity the Bessel index
$\nu$ is real. We apply the asymptotics of the Bessel functions
for large and positive arguments $\rho$ \cite{Erdelyi2},
\begin{eqnarray}
\label{bessel} J_\nu(\rho) &\sim& \frac{1}{\sqrt{2\pi\rho}}
\left[\exp\left(
i\rho-i\nu\frac{\pi}{2}-i\frac{\pi}{4}\right)\right.
\nonumber\\
& & \quad\quad\quad + \left.\exp\left(
-i\rho+i\nu\frac{\pi}{2}+i\frac{\pi}{4}\right)\right] \,.
\end{eqnarray}
We see that in the far field the light waves with real Bessel
indices are in a perfectly balanced superposition of incident and
emerging plane waves. In other words, the light is totally
reflected away from the interface of zero group velocity, similar
to the reflection of radio waves at the Earth's ionosphere
\cite{Jackson}.

A more interesting scenario appears in the second case when the
light is sufficiently blue-detuned to evoke an imaginary Bessel
index
\begin{equation}
\label{mu} \nu = i\mu \,.
\end{equation}
This regime can be reached by adjusting the gradient of the
control field that plays a decisive role in Eqs.\ (\ref{control})
and (\ref{catastrophe}). The smaller the gradient is the larger
is $a^2$. Therefore, a sufficiently small control-field gradient
gives rise to an imaginary Bessel index $\nu$, even within the
narrow transparency window in frequency $\omega=kc$. On the other
hand, as we will see at a later stage, the gradient should be as
large as possible for producing a maximal quantum effect. When
the Bessel index $\nu$ is imaginary the incident and the
reflected waves are not balanced anymore, as we see from the
asymptotics (\ref{bessel}) of the Bessel functions. Only a
fraction of the incident wave is reflected and the rest must be
transmitted somewhere. To find the transmitted wave, we focus on
the behavior of the Bessel functions for small arguments. We use
the first term in the power series \cite{Erdelyi2},
\begin{equation}
\label{power} J_{i\mu}(\zeta) \sim \frac{1}{(i\mu)!}
\left(\frac{\zeta}{2}\right)^{i\mu}\,,\quad \zeta=kz \,,\quad
\zeta\rightarrow 0 \,.
\end{equation}
Therefore, near the interface of zero group velocity the light
waves are proportional to
\begin{equation}
\zeta^{i\mu+1/2} = \sqrt{\zeta}\,\exp\left(i\mu\ln \zeta\right) =
\sqrt{\zeta}\,\exp(iS) \,.
\end{equation}
The logarithmic phase $S$ reduces dramatically the wave length
near $z=0$, because
\begin{equation}
\lambda = \frac{2\pi}{S_z} = \frac{2\pi}{\mu}\,z \,.
\end{equation}
The transmitted wave thus freezes in front of the interface of
zero group velocity.

We regard a process that creates an interface where waves separate
and develop a logarithmic phase singularity as a {\it wave
catastrophe}. The gravitational collapse of a star into a black
hole has created an horizon that cuts space into two disconnected
parts as well \cite{Misner}. Close to the event horizon, an
outside observer would see a similar behavior of waves
\cite{Brout}. All motion freezes near the horizon of the hole
which, therefore, has also been termed frozen star \cite{Thorne}.
In view of this analogy we regard the interface of zero group
velocity $\cal Z$ as the optical analog of a horizon.

\subsection{Modes}

Consider a superposition of sufficiently blue-detuned slow-light
waves (\ref{phi}) with imaginary Bessel indices (\ref{mu}). To
analyze the quantum effects of the horizon on polaritons, we must
turn the waves into modes (\ref{modes}), {\it i.e.}, we must
normalize the wave functions according to the scalar product
(\ref{scalar}). Waves (\ref{phi}) with different wave numbers $k$
are orthogonal to each other and possess a continuous spectrum.
Hence they should be normalized to delta functions. We employ the
ratio $k/k_0$ as the dimensionless mode index $q$ that occurs in
the scalar product (\ref{scalar}). The normalization factor is
entirely determined by the way in which the norm (\ref{scalar})
diverges to reach the $\delta$ singularity \cite{LL3}.
Consequently \cite{LL3}, we can ignore all finite, converging
contributions to the normalization integral given by Eqs.\
(\ref{scalar}), (\ref{catastrophe}), and (\ref{phi}),
\begin{eqnarray}
\label{integral} \left(\varphi_1,\varphi_2\right) & = & (\omega_1
+ \omega_2)  \int_0^\infty J_{\pm
i\mu_1}^*(k_1 z)\,J_{\pm i\mu_2}(k_2 z)\,z \nonumber\\
& & \quad\quad\quad\times\left(1 + \frac{a^2}{z^2}\right)\, dz \,.
\end{eqnarray}
The integral (\ref{integral}) diverges both at $z=0$ and at
$z=\infty$. We account for the two divergences seperately in the
integrals $I_0$ and $I_\infty$,
\begin{equation}
\label{sum} \left(\varphi_1,\varphi_2\right) = 2\omega \left(I_0 +
I_\infty\right) \,.
\end{equation}
Ignoring any convergent contributions to $I_0$, we cut off the
integral at some small $z_0$, regard $z(1+a^2/z^2)$ as $a^2/z$,
and use the asymptotics (\ref{power}) near the origin,
\noindent
\begin{eqnarray}
I_0 & = & \int_0^{z_0} J_{\pm i\mu_1}^*(k_1 z)\,J_{\pm i\mu_2}(k_2
z)\,\frac{a^2}{z}\,dz \nonumber\\
& = & \frac{a^2}{|(i\mu)!|^2} \int_0^{\zeta_0} \zeta^{\pm
i(\mu_2-\mu_1)}\, \frac{d\zeta}{\zeta} \nonumber\\
& = & {\rm sinh}(\pi\mu)\,\frac{a^2}{\mu\pi} \int_{-\infty}^{\ln
\zeta_0}\exp\left[\pm i (\mu_2-\mu_1)\xi\right]\,d\xi \nonumber\\
& = & {\rm sinh}(\pi\mu)\,\frac{a^2}{\mu}\,\delta(\mu_2-\mu_1)
\nonumber\\
& = & k^{-1}{\rm sinh}(\pi\mu)\,\delta(k_1-k_2) \,. \label{i0}
\end{eqnarray}
In the last step we have utilized that $\mu(\partial\mu/\partial
k)$ equals $a^2 k$. Furthermore, we have used Eq.\ 1.2.(6) of
Ref.\ \cite{Erdelyi1} for the gamma function
$x!\equiv\Gamma(x+1)$. Let us address the other integral,
\begin{eqnarray}
I_\infty & = & \int_{z_\infty}^\infty J_{\pm i\mu_1}^*(k_1
z)\,J_{\pm i\mu_2}(k_2
z)\,z\,dz \nonumber\\
& = & \frac{1}{2\pi k}\int_{z_\infty}^\infty \left[\exp\left(
-ik_1z \pm \mu_1\frac{\pi}{2} + i\frac{\pi}{4}\right)\right.
\nonumber\\
& & \quad\quad\quad\quad\left. + \exp\left(+ik_1z \mp
\mu_1\frac{\pi}{2} - i\frac{\pi}{4}\right)\right]
\nonumber \\
&& \quad\quad\quad \times \left[\exp\left(+ik_2z \pm
\mu_2\frac{\pi}{2} - i\frac{\pi}{4}\right)\right.
\nonumber\\
& & \quad\quad\quad\quad\left. + \exp\left(-ik_2z \mp
\mu_2\frac{\pi}{2} + i\frac{\pi}{4}\right)\right]\,dz
\nonumber\\
& = & \frac{1}{2\pi k}\int_{z_\infty}^\infty
\left[\exp\left(i(k_2-k_1)z \pm
(\mu_1+\mu_2)\frac{\pi}{2}\right)\right.
\nonumber\\
&& \left. \quad\quad\quad\quad
 + \exp\left(i(k_1-k_2)z \mp
(\mu_1+\mu_2)\frac{\pi}{2}\right) \right]\,dz
\nonumber\\
& = & k^{-1}{\rm cosh}(\pi\mu)\,\delta(k_1-k_2) \,. \label{inv}
\end{eqnarray}
Combining the two integrals (\ref{i0}) and (\ref{inv}) gives in
total (\ref{sum})
\begin{equation}
\left(\varphi_1,\varphi_2\right) = 2c\,e^{\pi\mu}\,\delta(k_1-k_2)
\,.
\end{equation}
Consequently, in the spatial region right from the horizon
$(z>0)$ the normalized wave functions are
\begin{equation}
\label{ur} u_R^\pm = \frac{\Theta(z)}{\sqrt{2c}}\,e^{-\pi\mu/2}\,
\sqrt{k_0 z}\,J_{\pm i\mu}(kz)\,e^{-i\omega t} \,.
\end{equation}
Left from the horizon $(z<0)$ we chose for convenience,
\begin{equation}
\label{ul} u_L^\pm(z) = u_R^\mp(-z)\,.
\end{equation}
The step function $\Theta$ in the definition (\ref{ur})
guarantees that the $u$ modes on different sides of the horizon
do not overlap and, therefore, they are automatically orthogonal
to each other. However, the $\pm$ degenerated waves on the same
side are not orthogonal. In fact, we obtain along similar lines
as in the normalization procedure,
\begin{equation}
\left(u_R^+,u_R^-\right) = e^{-\pi\mu}\,\delta(q_1-q_2) \,.
\end{equation}
Yet we construct easily the orthogonal partners $w^\pm$ to the
$u^\mp$ waves,
\begin{eqnarray}
\label{w} w_R^\pm & = & \frac{1}{\sqrt{1-e^{-2\pi\mu}}} \,\left(
u_R^\pm - e^{-\pi\mu}\,u_R^\mp\right) \,,\nonumber\\
w_L^\pm(z) & = &w_R^\mp(-z) \,.
\end{eqnarray}
We chose as the orthonormal basis in the mode expansion
(\ref{modes})
\begin{equation}
\label{set} w_R \equiv w_R^+ \,,\quad u_R \equiv u_R^- \,,\quad
w_L \equiv w_L^- \,,\quad u_L \equiv u_L^+ \,.
\end{equation}
Finally, to find an interpretation of the $w$ modes, we use the
asymptotics (\ref{bessel}) of the Bessel functions and get for
$|z|\rightarrow\infty$ on the appropriate sides of the horizon,
\begin{eqnarray}
w_R & \sim & \left(\frac{1 - e^{-2\pi\mu}}{4\pi c\,
\omega/\omega_0}\right)^{1/2} \exp\left(+i\frac{\omega}{c}\,z -
i\omega t - i\frac{\pi}{4}\right)
\,,\nonumber\\
w_L  & \sim & \left(\frac{1 - e^{-2\pi\mu}}{4\pi c\,
\omega/\omega_0}\right)^{1/2} \exp\left(-i\frac{\omega}{c}\,z -
i\omega t - i\frac{\pi}{4}\right) \,. \label{hankel}
\end{eqnarray}
The asymptotics (\ref{hankel}) shows that the $w$ modes turn into
plane waves propagating away from the horizon. In other words,
the $w$ modes are the ones that reach an external photon detector.

\subsection{Analyticity}

The stationary modes (\ref{set}) of catastrophic slow light are
severely non-analytic. The modes vanish on either the left or the
right side of the horizon with a characteristic essential
singularity as a precursor. Waves near the event horizon of a
black hole suffer a similar fate \cite{Brout}. Seen from an
outside observer, the waves freeze near the Schwarzschild radius
$r_s$ with an essential singularity of the type $(r-r_s)^{i\mu}$
where $\mu=2r_s\omega/c$ \cite{Brout}. Yet an observer falling
into the hole would see little difference in waves near the
horizon and could pass the point of no return without noticing.
Like the inward-falling observer, the quantum vacuum flows towards
the central singularity of the hole and, similarly, the horizon
should not be a special place for the vacuum either. In
mathematical terms, the wave function of the vacuum is analytic
\cite{Brout,Unruh}. Consequently, the modes seen by the outside
observer must not be in their vacuum states. In fact, they carry
the quanta of Hawking radiation \cite{Hawking}. The history of
the hole formation during a gravitational collapse turns out to be
responsible for the analyticity of the vacuum \cite{Brout,Unruh}.
Inspired by the analogy between a black hole and our slow-light
catastrophe, let as consider the history of our horizon.

Suppose that the group index $\alpha$ was initially a largely
uniform $\alpha_0$. Then, by tuning the control field, the group
index develops a quadratic singularity, for example as $\alpha =
a^2(t)/(z^2 + b^2(t))$ with $a^2(-\infty)\rightarrow\infty$,
$b^2(-\infty)\rightarrow\infty$,
$a^2(-\infty)/b^2(-\infty)\rightarrow\alpha_0$, and finally
$b^2(+\infty)\rightarrow 0$. The details of the process do not
matter. Immediately after a time $t_c$ the group index will
possess the quadratic singularity that we are studying, creating a
slow-light catastrophe. Given a uniform group index as the
initial condition and no slow-light injected, the polariton
vacuum occupies initially packets of plane waves that we can sort
into right- or left-traveling waves,
\begin{eqnarray}
\label{initial} \varphi_\pm & = & \int A_\pm
(\omega)\,\exp\left(\pm i\frac{z}{c}\,\sqrt{\omega^2 + \alpha_0
(\omega^2 - \omega_0^2)} - i\omega t \right)\nonumber\\
& & \quad\times d\omega \,.
\end{eqnarray}
Regarded as a function of complex $z$, the $\varphi_\pm$ waves are
analytic in the upper $(+)$ or lower $(-)$ half plane,
respectively, because here the integral ({\ref{initial})
converges. When the control field creates a horizon the vacuum
modes must follow the wave equation (\ref{wave}). We assume that
$\alpha$ is analytic apart from poles. Consider closed contour
integrals in either one of the half planes. We obtain from the
wave equation (\ref{wave}),
\begin{equation}
\label{oint}
\partial_t \oint (1 + \alpha) \varphi_t \,dz = c^2 \oint
\varphi_{zz}\,dz - \omega_0^2 \oint \alpha\varphi\,dz \,.
\end{equation}
Due to the analyticity of the initial wave packets
(\ref{initial}) $\oint\varphi\,dz$ and $\oint\varphi_t\,dz$ were
initially zero. Equation (\ref{oint}) indicates that both
integrals remain zero, as long as $\alpha$ is analytic. At single
poles of $\alpha$ we get $\varphi_t = -\omega_0^2\varphi$, which
cannot generate a singularity. Higher poles of $\alpha$ do not
contribute to the closed contour integrals. Consequently, the
vacuum wave functions are always analytic in $z$.

\vspace*{1mm}
\begin{figure}[htbp]
\begin{center}
\includegraphics[width=20.5pc]{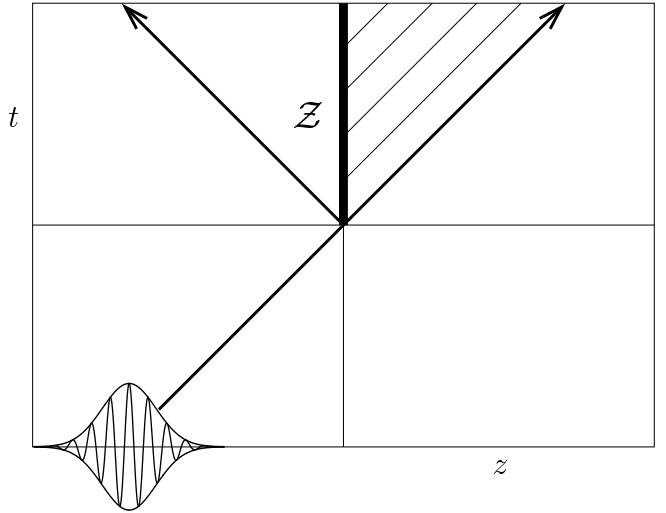}
\vspace*{2mm} \caption{Space-time diagram of a slow-light
catastrophe. The figure illustrates the fate of a wave packet
$\varphi(t,z)$ that experiences the formation of the horizon
${\cal Z}$. Initially, the packet oscillates with positive
frequencies in time $t$ and propagates from the left to the right
in space $z$. The horizon cannot generate negative frequencies in
the reflected light, apart from a brief burst that we neglect. On
the left side of ${\cal Z}$ we thus regard $\varphi(t,z)$ as
analytic in $t$ on the lower half of the complex $t$ plane.
Furthermore, $\varphi(t,z)$ is analytic in $z$ on the upper half
plane throughout the history of the wave packet, because the
process (\ref{wave}) conserves analyticity. Yet $\varphi(t,z)$ is
not analytic in $t$ on the other side of the horizon, as the
solution (\ref{solution}) indicates. Here waves with negative
frequencies are continuously peeling away from the horizon,
corresponding to a stationary creation of slow-light quanta.}
\label{figure2}
\end{center}
\end{figure}

Consider the analytic properties of the vacuum waves with respect
to time. Picture a wave with positive frequencies incident from
the left, see Fig.\ 2. After the catastrophe a part of the wave
may freeze at the horizon and the rest is reflected. If the wave
happens to arrive during the formation of the horizon a brief
burst of light with negative frequencies may be generated.
However, in the stationary regime we are interested in, the
reflected light contains always positive frequencies. Therefore
we regard the wave function $\varphi(t,z)$ on the left side of
the horizon as analytic in $t$ on the lower complex plane. Given
an analytic signal $\varphi(t)$ at some $z$ we can propagate it
in space according to the wave equation (\ref{wave}) without loss
in analyticity, but we cannot pass the horizon, because here
$\alpha$ is singular. It is therefore conceivable that beyond the
horizon $\varphi(t)$ is not analytic anymore. In other words,
$\varphi(t)$ may contain negative frequencies. In fact, we show
in the next subsection that negative frequencies in time are
unavoidable for not running into conflict with the analyticity in
the spatial coordinate $z$. Waves propagating to the right are
analytic in $z$ on the upper half plane and they must have
originated from a wave incident from the left. The analytic
properties of the vacuum waves in space are thus connected with
their analytic properties in time. Analyticity in the upper half
of the $z$ plane is linked to analyticity in $t$ on the left side
and, using similar arguments, analyticity in the lower half of
the $z$ plane goes hand in hand with analyticity in $t$ on the
right side. We utilize the analytic properties of the vacuum
waves in space and time as a marker for distinguishing the vacuum
modes.

\subsection{Combinations}

The vacuum states of slow-light polaritons are characterized by
analytic wave functions in $z$. Therefore, to describe the vacuum
after the formation of the horizon, we should construct
orthonormal combinations $v$ of the non-analytic $u$ and $w$
modes that are analytic on either the upper or the lower half of
the complex $z$ plane at some arbitrary time $t_0$. The solution
is
\begin{eqnarray}
v_R & = & {\textstyle \frac{1}{2}}{\rm sech}(\pi\mu) w_L - iw_R
-ie^{-2i\omega t_0} {\textstyle\frac{1}{2}} {\rm sech} (\pi\mu)
w_R^* \,,\nonumber\\
v_R^\perp & = & \frac{1}{\sqrt{1 + e^{2\pi\mu}}}\,\left(u_R +
ie^{\pi\mu} u_L\right)
\,,\nonumber\\
v_L & = & {\textstyle \frac{1}{2}}{\rm sech}(\pi\mu) w_R - iw_L
-ie^{-2i\omega t_0} {\textstyle\frac{1}{2}} {\rm sech}
(\pi\mu) w_L^* \,,\nonumber\\
v_L^\perp & = & \frac{1}{\sqrt{1 + e^{2\pi\mu}}}\,\left(u_L +
ie^{\pi\mu} u_R\right)\,. \label{solution}
\end{eqnarray}
Because the $u$ and $w$ modes are orthonormal with respect to the
scalar product (\ref{scalar}), one can easily verify that the $v$
modes form an orthonormal set as well.

The modes (\ref{solution}) are given on the real $z$ axis and are
subject to analytic continuation. We prove that the $v_R$ and
$v_R^\perp$ modes are analytic on the upper half plane and that
the $v_L$ and $v_L^\perp$ modes are analytic on the lower half
plane. First, we use the definitions (\ref{w}) and (\ref{set}),
and write $v_R$ and $v_L$ in the form
\begin{eqnarray}
\lefteqn{2{\rm cosh}(\pi\mu)\sqrt{1 - e^{-2\pi\mu}}\, v_R}
\nonumber \\
& & = u_L^- -ie^{+\pi\mu}\left((1 + e^{-2\pi\mu})u_R^+ -
e^{-2\pi\mu}{u_R^-}^* e^{-2i\omega t_0}\right) \nonumber\\
& & \quad - \,\,e^{-\pi\mu}\left[u_L^+ - ie^{-\pi\mu}\left((1 +
e^{+2\pi\mu})u_R^- \right.\right.\nonumber\\
& & \quad\quad\left.\left. - e^{+2\pi\mu}{u_R^+}^* e^{-2i\omega
t_0}\right)\right]
\nonumber\\
\lefteqn{2{\rm cosh}(\pi\mu)\sqrt{1 - e^{-2\pi\mu}}\, v_L}
\nonumber \\
& & = u_R^+ -ie^{+\pi\mu}\left((1 + e^{-2\pi\mu})u_L^- -
e^{-2\pi\mu}{u_L^+}^* e^{-2i\omega t_0}\right) \nonumber\\
& & \quad - \,\,e^{-\pi\mu}\left[u_R^- - ie^{-\pi\mu}\left((1 +
e^{+2\pi\mu})u_L^+ \right.\right.\nonumber\\
& & \quad\quad\left.\left. - e^{+2\pi\mu}{u_L^-}^* e^{-2i\omega
t_0}\right)\right] \,.
\end{eqnarray}
Then we show that the combinations
\begin{eqnarray}
u_L^- - ie^{+\pi\mu} u_R^+ \,&,&\, u_L^- - ie^{+\pi\mu}
{u_R^-}^* e^{-2i\omega t_0} \,,\nonumber\\
u_L^+ - ie^{-\pi\mu} u_R^- \,&,&\, u_L^+ - ie^{-\pi\mu} {u_R^+}^*
e^{-2i\omega t_0} \label{up}
\end{eqnarray}
are analytic on the upper half plane and that the corresponding
combinations
\begin{eqnarray}
u_R^+ - ie^{+\pi\mu} u_L^- \,&,&\, u_R^+ - ie^{+\pi\mu}
{u_L^+}^* e^{-2i\omega t_0} \,,\nonumber\\
u_R^- - ie^{-\pi\mu} u_L^+ \,&,&\, u_R^- - ie^{-\pi\mu} {u_L^-}^*
e^{-2i\omega t_0} \label{low}
\end{eqnarray}
are analytic on the lower half plane. The analyticity of the modes
(\ref{solution}) follows from the analytic properties of the
combinations (\ref{up}) and (\ref{low}). Here it is sufficient to
focus on the vicinity of the origin where the left and right
modes are connected. As a consequence of Eqs.\ (\ref{power}),
(\ref{ur}) and (\ref{ul}), we get in terms of $\zeta = kz$,
\begin{eqnarray}
u_R^\pm &\sim&
\frac{\Theta(z)}{\sqrt{2c\,\omega/\omega_0}}\,e^{-\mu\pi/2}\,
\frac{2^{\mp i\mu}}{(\pm i\mu)!}\,e^{-i\omega t}\,\zeta^{\pm i\mu
+ 1/2} \,, \nonumber\\
u_L^\pm(\zeta) & = & u_R^\mp(-\zeta)\,.
\end{eqnarray}
Consider the analytic properties of $\zeta^{i\mu + 1/2}$ for an
arbitrary real $\mu$. We indicate with a suffix $\pm$ whether
$\zeta^{i\mu + 1/2}$ should be regarded as analytic on the upper
$(+)$ or on the lower $(-)$ half plane, respectively. With this
notation we get
\begin{eqnarray}
\zeta_\pm^{i\mu + 1/2} & = & \Theta(\zeta)\,\zeta^{i\mu + 1/2} +
\Theta(-\zeta)\,\zeta_\pm^{i\mu + 1/2} \nonumber\\
& = & \Theta(\zeta)\,\zeta^{i\mu + 1/2} +
\Theta(-\zeta)\,(-1)_\pm^{i\mu + 1/2}\,(-\zeta)^{i\mu + 1/2}
\nonumber\\
& = & \Theta(\zeta)\,\zeta^{i\mu + 1/2} + \Theta(-\zeta)\, e^{\pm
i\pi(i\mu + 1/2)}\,(-\zeta)^{i\mu + 1/2}
\nonumber\\
& = & \Theta(\zeta)\,\zeta^{i\mu + 1/2} \pm i
e^{\mp\pi\mu}\,\Theta(-\zeta)\,(-\zeta)^{i\mu + 1/2} \,.
\end{eqnarray}
This relation proves the analyticity of the combinations
(\ref{up}) and (\ref{low}) and, as a result, the analyticity of
the modes (\ref{solution}).

However, is the set of modes (\ref{solution}) unique? In
principle, we could perform linear canonical transformations that
convert the modes (\ref{solution}) into a new set, yet such
transformations are severely restricted by the required
analyticity in space and time. For example, we can construct
superpositions of the $v_R$ and $v_R^\perp$ modes to form the new
modes $v_R\cos\theta + e^{i\gamma}v_R^\perp\sin\theta$ and
$v_R^\perp \cos\theta - e^{-i\gamma}v_R\sin\theta$, analogous to
the mode transformation of a beam splitter \cite{Splitter}. Or we
may combine $v_R$ with $v_L^*$ in the Bogoliubov transformation
$v_R\cosh\xi + e^{i\gamma}v_L^*\sinh\xi$ and $v_R^\perp\cosh\xi +
e^{i\gamma}v_L^*\sinh\xi$, analogous to a parametric amplifier
\cite{Amplifier}. However, only transformations of this type
maintain the analyticity on one of the half planes of complex
$z$. We are not allowed to combine $v_R$ with ${v_R^\perp}^*$ or
$v_R$ with $v_L$. Yet the possible mode transformations are
further restricted: a Bogoliubov transformation $v_R\cosh\xi +
e^{i\phi}v_L^*\sinh\xi$ would generate negative frequencies on the
left side of the horizon. We have argued that this must not
happen. Similar arguments apply to all other Bogoliubov
transformations. Consequently, our modes are uniquely defined up
to superpositions, but such transformations do not change the
vacuum state \cite{Splitter}. Therefore, the $v$ modes are indeed
the vacuum modes.

\subsection{Radiation}

As a consequence of a slow-light catastrophe, the polariton field
is decomposed into two different sets of modes. The $v$ modes
contain the polariton vacuum, whereas the $u$ and $w$ modes guide
the detectable quanta,
\begin{eqnarray}
\hat{\varphi} & = & \int \Big(\hat{a}_R w_R + \hat{a}_{R\perp}
u_R + \hat{a}_L w_L + \hat{a}_{L\perp} u_L + {\rm H.c.} \Big)\, dq
\,,\nonumber\\
& = & \int \Big(\hat{b}_R v_R + \hat{b}_{R\perp} v^\perp_R +
\hat{b}_L v_L + \hat{b}_{L\perp} v^\perp_L + {\rm H.c.} \Big)\, dq
\,.
\end{eqnarray}
The $\hat{a}$ operators are the annihilation operators of the
detector modes and the $\hat{b}$ operators refer to the vacuum
modes. H.c.\ denotes the Hermitian conjugate and the $q$ are the
mode indices $k/k_0$. Notice that the vacuum modes contain both
positive and negative frequency components, because the set
(\ref{solution}) involves complex conjugate $w$ modes. Therefore
we expect that the corresponding operator transformations combine
annihilation with creation operators. This is the decisive sign
of pair creation, similar to the production of photon pairs in
parametric downconversion \cite{MandelWolf}. We represent
$\hat{a}_R$ as $(w_R,\hat{\varphi})$, $\hat{a}_{R\perp}$ as
$(u_R,\hat{\varphi})$ et cetera, use the normalization
(\ref{norm}) and the properties of the scalar product
(\ref{scalar}), and arrive at the Bogoliubov transformations
\cite{Birrell}
\begin{eqnarray}
\hat{a}_R & = & {\textstyle \frac{1}{2}}{\rm sech}(\pi\mu)
\hat{b}_L - i\hat{b}_R +ie^{2i\omega t_0} {\textstyle\frac{1}{2}}
{\rm sech} (\pi\mu) \hat{b}_R^\dagger \,,\nonumber\\
\hat{a}_{R\perp} & = & \frac{1}{\sqrt{1 +
e^{2\pi\mu}}}\,\left(\hat{b}_{R\perp} + ie^{\pi\mu}
\hat{b}_{L\perp}\right)
\end{eqnarray}
and at analogous relations for the $\hat{a}_L$ and
$\hat{a}_{L\perp}$. Without initial probe light injected, the
dynamically formed slow-light catastrophe will cause spontaneous
radiation of probe polaritons at a constant rate, because we
obtain for $v$-mode vacua
\begin{equation}
\langle\hat{a}_R^\dagger(q_1)\hat{a}_R(q_2)\rangle =
\langle\hat{a}_L^\dagger(q_1)\hat{a}_L(q_2)\rangle =
\bar{n}\,\delta(q_1 - q_2)
\end{equation}
with the average particle number
\begin{equation}
\label{leo} \bar{n} = \frac{1}{\left(e^{\pi\mu} +
e^{-\pi\mu}\right)^2} \,.
\end{equation}
The radiation energy will be taken from the control beam. The
initial formation of the horizon is a time-dependent process
that, therefore, transfers energy to the polariton field. Yet, in
addition to an initial brief burst of energy, the control beam
creates a wave catastrophe that would force polaritons into a
state they cannot occupy. The frustrated polariton field reacts in
attempting to alter the parabolic profile of the control
intensity. This process takes energy away from the control beam
and allows the creation of polarition pairs. Pair production
continues as long as the control beam is not significantly
depleted. A running wave of control light will produce a steady
flow of slow-light quanta.

The slow-light catastrophe generates a maximal particle number per
mode (\ref{leo}) of ${1}/{4}$, which is quite substantial,
considering the fact that bright sunlight with a radiation
temperature of $6\cdot 10^3 {\rm K}$ carries a mere $0.01$
photons per mode in the optical range of the Planck spectrum.
However, the photon number (\ref{leo}) is sharply peaked as a
function of $\mu$ and, in any case, our pair-production mechanism
is restricted to the narrow frequency window of EIT \cite{EIT}.
To maximize the generated quantum radiation, one should create a
situation where $\mu$ is near zero over an as large as possible
spectral range. In terms of the experimental parameters, we get
in the transparency window near $\omega_0$,
\begin{eqnarray}
\label{parameters} \mu & = &
\frac{1}{2}\left(\frac{\delta}{\delta_0} - 1\right)^{1/2} ,\quad
\delta = \frac{\omega - \omega_0}{\omega_0} \,,\nonumber\\
\delta_0 & = & \frac{c^2}{8a^2\omega_0^2} =
\frac{1}{32\pi^2}\left(\frac{\lambda_0}{a}\right)^2 \,.
\end{eqnarray}
Pair production occurs on the blue side of the critical detuning
$\delta_0$ (for $\delta>\delta_0$) where $\delta_0$ also
determines the width of the spectrum (\ref{leo}). The smaller the
scale $a$ of the group-index profile (\ref{catastrophe}) is, the
larger is the critical detuning (\ref{parameters}) and the wider
is the spectrum (\ref{leo}) of particle production. Equations
(\ref{control}) and (\ref{group}) indicate that the scale $a$ is
small for a steep group-velocity profile created by a large
control-field gradient. Gravitational black holes show a similar
behavior \cite{Hawking}. The smaller the hole is, the larger is
the gravity gradient at the horizon and the stronger is the
Hawking radiation \cite{Hawking} generated. Returning to our
case, the parameter-dependence of the particle-number spectrum
(\ref{leo}) underlines the crucial role of a spatially varying
group velocity in creating a quantum catastrophe. A mere zero of
the group velocity would not suffice to produce a measurable
radiation. The control-field gradient matters.

Close to the horizon the susceptibility of slow light diverges.
Yet Nature tends to prevent infinite susceptibilities: Instead of
responding infinitely strongly, optical media become absorptive or
non-linear. Considering gravitational black holes, Nature could
prevent the existence of true event horizons as well. Here waves
are supposed to shrink in wave lengths beyond the Planck scale
where the physics is unknown. Hawking radiation seems to stem
from these extremely shortened waves. This trans-Planckian
problem \cite{TransPlanckProblem} was analyzed in theoretical toy
models of sonic black holes in moving fluids
\cite{Sonic,TransPlanck}. Here the inter-atomic distance provides
a natural cut-off for extremely shortened sound waves.
Nevertheless \cite{TransPlanck}, the mere threat of a horizon
seems to be sufficient for generating Hawking sound.

Returning to slow light, an EIT medium in linear response is
transparent within a narrow spectral window around the resonance
frequency $\omega_0$. The spectral width of the transparency
window is proportional to the intensity of the control beam
\cite{EIT}. Therefore, slow-light waves oscillating at frequencies
different from $\omega_0$ are absorbed near a node of the control
field, unless the medium becomes non-linear. We show in Appendix
A that the non-linearity of the EIT medium depends on the ratio
of the probe and control intensities. According to the linear
optics of slow light discussed in the body of this paper, the
probe intensity is proportional to the distance $z$ from the
horizon, as long as $kz$ is small, whereas the control intensity
grows quadratically. Consequently, at a certain distance $z_0$
both fields are comparable in strength. Here slow light leaves the
regime of linear response. In a semi-classical concept of light
\cite{Paul,Bachor} quantum fluctuations are small perturbations of
the classical amplitude and are subject to a linear theory. If
the vacuum state of light is classically unstable, photons are
created spontaneously. For example, in parametric downconversion
\cite{MandelWolf} quantum fluctuations are amplified \cite{Paul},
generating photon pairs. The instability of the linear optics
near a slow-light horizon may provide the microscopic mechanism
for the pair production we have predicted phenomenologically.
However, the non-linear effects of the EIT medium are required to
dominate at a stage where the absorption is still small. We show
in Appendix B that the probe intensity is proportional to the
detuning $\delta_0$ given by Eq.\ (\ref{parameters}) in terms of
the characteristic scale (\ref{catastrophe}) of the group index
profile (\ref{control}). Therefore, the ratio of the control and
probe intensities does not depend on $a$, and, consequently, the
non-linearity distance $z_0$ is independent of the control-field
gradient. In order to avoid absorption the control should be
strong enough at $z_0$ which requires a steep field gradient.
Using the experimental parameters of Refs.\ \cite{Liu,Dutton}, the
Rabi frequency \cite{MandelWolf} of the control field should grow
at least by $10 {\rm MHz}$ per wave length $\lambda_0$ in distance
from the horizon, as we show in Appendix B. In this case the scale
$a=5\times10^3\lambda_0$ and the critical detuning
$\delta_0=10^{-10}$. Appendix B indicates that millions of
photons are generated per second, amounting to a gentle glow
perhaps visible with the naked eye.

\section{Summary}

Tuning the control field towards a parabolic intensity profile
causes a catastrophic situation for slow-light polaritons. In
turn, the polariton field sets out to deplete the control beam,
in an attempt to alter the intensity profile that has caused the
wave catastrophe in the first place, yet in vain. The control beam
continuously replenishes the parabolic intensity profile, driving
a stationary production of polariton pairs. The two polaritons of
each pair are created on opposite sides near the horizon, they
depart at a snail's pace, accelerate gradually and emerge as
detectable photons. The Hawking radiation of a black hole
\cite{Hawking} follows a similar scenario \cite{Brout}. Here the
gravitational collapse \cite{Misner} has triggered a quantum
catastrophe at the event horizon, causing pair creation lasting
as long as the hole possesses gravitational energy
\cite{Brout,Hawking}. One particle of each pair falls into the
black hole, whereas the other escapes into space and appears as
thermal radiation \cite{Hawking}. In our case, and in contrast to
gravitational holes, one can explore the other side beyond the
horizon and, for example, measure the correlations of the
generated photon pairs. Both cases are triggered by catastrophic
events with lasting consequences.

The quantum radiation of a slow-light catastrophe resembles
Hawking radiation but also exhibits some interesting differences.
The emitted spectrum (\ref{leo}) is not Planckian, whereas a
black hole of Schwarzschild radius $r_s$ appears as a black-body
radiator with temperature $\hbar c/(4\pi r_s)$ \cite{Hawking}.
The differences between the two spectra can be traced back to two
different classes of wave catastrophes. In both cases, waves
freeze at an horizon in the form $\zeta^p$ with an exponent $i\mu
+ {1}/{2}$ for slow-light media but with an exponent $i\mu$ for
black holes where $\mu = 2\pi r_s\, \omega/c$ \cite{Brout}. Note
that Unruh's effect \cite{Unruh} of radiation seen by an
accelerated observer is of Hawking-class as well \cite{Brout} and
so are most of the proposed artificial black holes
\cite{LPliten,Sonic,TransPlanck,Reznik,Chapline,Volovik}.
Remarkably, Schwinger's pair production of charged particles in
electrostatic fields \cite{Schwinger} is accompanied by a subtle
wave catastrophe of exponent $i\mu - {1}/{2}$ \cite{Brout} and
leads to a Boltzmannian spectrum $\bar{n}=\exp(-2\pi\mu)$. All
three catastrophes agree in the limit of large $\mu$ but deviate
significantly in the regime of maximal particle production where
$\mu$ is small. It might be interesting to find out whether more
than three types of quantum catastrophes can occur.

\section*{Acknowledgements}

I am very grateful to Sir Michael Berry, Lene Vestergaard Hau,
Malcolm Dunn, Tamas Kiss, Patrik \"Ohberg, Renaud Parentani, Paul
Piwnicki, Paul Sheldon, and Matt Visser for inspiring
conversations. In particular, Sir Michael \cite{Berry} raised the
question whether wave singularities exhibit interesting quantum
effects. The paper was supported by the ESF Programme Cosmology
in the Laboratory.

\section*{Appendix A: Dark-state dynamics}

In this appendix we consider the microscopic theory of the atoms
constituting an EIT medium. We derive the wave equation
(\ref{wave}) as the linear-response limit of the non-linear
dynamics of slow light. Assume that the EIT medium consists of
$n_A$ identical atoms per unit volume, each one equipped with
three levels interacting near-resonantly with the probe and
control fields respectively, see Fig.\ \ref{levels}.
\begin{figure}
\begin{center}
\includegraphics[width=9cm]{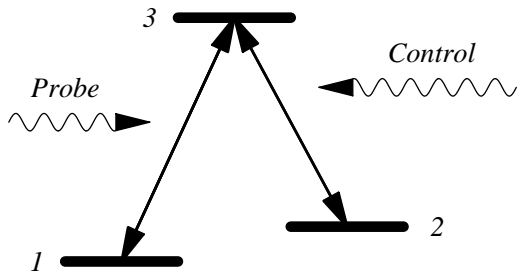}
\caption{Three-level atom in a regime of
Electromagnetically-Induced Transparency. The control beam
couples the levels 2 and 3, which influences strongly the optical
properties of the atom for a weaker probe beam tuned to the
transition 1$\leftrightarrow$3.} \label{levels}
\end{center}
\end{figure}
We treat the light as a classical electromagnetic field. An atom
is characterized by the energy-level differences
$\hbar\omega_{12}$ and $\hbar\omega_{23}$ with $\omega_{12} +
\omega_{23} = \omega_{13} \equiv \omega_0$. Typically, the
transition frequencies $\omega_{13}$ and $\omega_{23}$ are in the
optical range of the spectrum or in the near infrared
($10^{15}{\rm Hz}$), whereas the frequency $\omega_{12}$ is much
lower ($10^{9}{\rm Hz}$). The atom is subject to fast relaxation
mechanisms ($10^{6}{\rm Hz}$) that transport atomic excitations
from the $|\,3\,\rangle$ state down to $|\,1\,\rangle$ and from
$|\,3\,\rangle$ to $|\,2\,\rangle$, mainly caused by spontaneous
emission. Hardly any excitations move from  $|\,2\,\rangle$ to
$|\,1\,\rangle$, because the spontaneous emission rate is
proportional to the cube of the frequency \cite{Loudon}. Here the
relaxation may be dominated by other processes, for instance by
spin-exchanging collisions. Without relaxation the dynamics of
the atom is governed by the Hamiltonian
\begin{equation}
\label{h} \hat{H} = \left[
\begin{array}{ccc}
0 & 0 & -\frac{1}{2}\,\kappa_{13}\,E_p^{(-)} \\[3mm]
0 & \hbar\omega_{12} & -\frac{1}{2}\,\kappa_{23}\,E_c^{(-)}\\[3mm]
-\frac{1}{2}\,\kappa_{13}\,E_p^{(+)} &
-\frac{1}{2}\,\kappa_{23}\,E_c^{(+)} & \hbar\omega_{13}
\end{array}
\right] \,.
\end{equation}
The Hamiltonian represents the atomic level structure and
describes the dipole interaction with light, considering here only
the positive/negative frequency components $E_p^{(\pm)}$ and
$E_c^{(\pm)}$ that match approximately the level structure. The
$E_p$ and $E_c$ fields are the probe and control light
respectively and are given in SI units. We describe relaxation
phenomenologically by the transition processes
\begin{equation}
\label{relax} \hat{A}_1 = |\,1\,\rangle\,\langle\,3\,|\,\,,\quad
\hat{A}_2 = |\,2\,\rangle\,\langle\,3\,|\,\,,
\end{equation}
occurring at the rates $\gamma_1$ and $\gamma_2$, typically a few
$10^{6}\,{\rm Hz}$. The density matrix of the atom, $\hat{\rho}$,
evolves according to the master equation \cite{Gardiner}
\begin{eqnarray}
\label{master} \frac{d\hat{\rho}}{d t} & = &
\frac{i}{\hbar}\,[\hat{\rho},\hat{H}]\nonumber\\
& & - \sum_l \gamma_l\left(\hat{A}_l^\dagger\hat{A}_l\hat{\rho} -
2\hat{A}_l\hat{\rho}\hat{A}_l^\dagger +
\hat{\rho}\hat{A}_l^\dagger\hat{A}_l \right) \,.
\end{eqnarray}
It is advantageous to represent the light fields in terms of Rabi
frequencies
\begin{equation}
\label{rabi} \Omega_c\, e^{-i\omega_c t} =
\frac{\kappa_{23}}{\hbar}\, E_c^{(+)} \,,\quad \Omega_p\,
e^{-i\omega_0 t} = \frac{\kappa_{13}}{\hbar}\, E_p^{(+)} \,,
\end{equation}
defined here with respect to the atomic transition frequencies
$\omega_c=\omega_{23}$ and $\omega_0=\omega_{13}$. In the absence
of relaxation, an atom would oscillate between the ground and the
excited state with frequency $\Omega$ (Rabi flopping
\cite{MandelWolf}). On the other hand, relaxation leads to a
stationary state where the atomic dipoles follow the fields.

Assume that the control beam is in exact resonance $\omega_c$ and
that the probe light is monochromatic with a small detuning
$\omega-\omega_0$. Furthermore, the Rabi frequency of the control
beam shall dominate all relevant time scales,
\begin{equation}
\label{regime} |\,\Omega_c| \gg
|\,\Omega_p|\,,\,\gamma_1\,,\,\gamma_2\,,|\,\omega-\omega_c| \,.
\end{equation}
In this limit the stationary state of the atomic evolution
(\ref{master}) turns out to approach the pure state
\begin{eqnarray}
\hat{\rho} & = & |\,\psi_0\,\rangle \langle \,\psi_0\,|
\nonumber\\[1mm]
|\,\psi_0\,\rangle & = & \hat{U}_0\, N_0 \left(|\,1\,\rangle -
\frac{\Omega_p}{\Omega_c}\,|\,2\,\rangle +
\frac{2(\omega-\omega_0)}{|\,\Omega_c|^2}\,\Omega_p\,|\,3\,\rangle
\right) \label{psi0}
\end{eqnarray}
that is called a dark state \cite{EIT}. Here we have separated the
rapid oscillations of the atom at the optical transition
frequencies from the slower atomic dynamics,
\begin{equation}
\hat{U_0}  =  \left[
\begin{array}{ccc}
1 & 0 & 0 \\
0 & e^{-i\omega_{12}t} & 0\\
0 & 0 & e^{-i\omega_0t}
\end{array}\right]
\,.
\end{equation}
Suppose that a dominant and monochromatic control beam has, after
relaxation, prepared the atom in the stationary state
(\ref{psi0}). How will the atom evolve when the control and probe
strengths vary? First we show that the atom remains in a pure
state, as long as the $|\,3\,\rangle$ component is small,
\begin{equation}
\label{condition} \rho_{33} = \langle \, 3\, |\,\hat{\rho}\,|\, 3
\,\rangle \ll 1 \,.
\end{equation}
Consider the statistical purity ${\rm tr}\{\hat{\rho}^2\}$. A
quantum system is in a pure state if and only if the purity is
unity \cite{Gardiner}. We apply the master equation
(\ref{master}) and see that the purity does not change
significantly,
\begin{eqnarray}
d {\rm tr}\{\hat{\rho}^2\} & = & 2 {\rm tr}
\{\hat{\rho}\,d\hat{\rho}\} \nonumber\\
& = & 4\left[\gamma_1(1-\rho_{11}) + \gamma_2(1-\rho_{22})
\right] \rho_{33}\, d t \,,
\end{eqnarray}
once the atom has occupied a pure state with sparsely populated
level $|\,3\,\rangle$. Consequently, we can describe the state of
the atom by a vector $|\,\psi\,\rangle$.

Suppose that the control and the probe strengths vary. How does a
dark state follow the light? In the case (\ref{condition}) the
state vector is dominated by its components in the subspace
spanned by the two lower levels $|\,1\,\rangle$ and
$|\,2\,\rangle$. If we find a vector $|\,\psi\,\rangle$ that
describes correctly the dynamics (\ref{master}) in this subspace,
the third component $\langle\,3\,|\,\psi\,\rangle$ must be
correct as well, to leading order in $\rho_{33}$. The lower ranks
enslave the top level. Since the relaxation processes
(\ref{relax}) do not operate within the lower subspace, we can
ignore dissipation entirely, to find the dominant state of the
atom.  We write down the state vector
\begin{equation}
\label{dark} |\,\psi\,\rangle = \hat{U}_0 \, N
\left(|\,1\,\rangle - \frac{\Omega_p}{\Omega_c}\,|\,2\,\rangle +
\frac{2N_0^2}{\Omega_c^*}\,i\partial_t\,\frac{\Omega_p}{\Omega_c}
\,|\,3\,\rangle \right)
\end{equation}
with the abbreviations
\begin{eqnarray}
\frac{\Omega_p}{\Omega_c} &=&
\left|\frac{\Omega_p}{\Omega_c}\right|\,e^{i\theta}\,,
\nonumber\\[2mm]
N_0 &=& \left(1 +
\frac{|\,\Omega_p|^2}{|\,\Omega_c|^2}\right)^{-1/2} \,,
\nonumber\\[2mm]
N &=&
N_0\,\exp\left(-i\int\frac{|\,\Omega_p|^2\,d\theta}{|\,\Omega_p|^2
+ |\,\Omega_c|^2}\right)\,.
\end{eqnarray}
In a stationary regime under the condition (\ref{regime}) the
vector (\ref{dark}) agrees with the dark state (\ref{psi0}). We
see from the properties
\begin{equation}
\partial_t N = -N N_0^2\,\frac{\Omega_p^*}{\Omega_c^*}\,\partial_t\,
\frac{\Omega_p}{\Omega_c} \,,\quad
\partial_t N\,\frac{\Omega_p}{\Omega_c} = N N_0^2\,\partial_t\,
\frac{\Omega_p}{\Omega_c}
\end{equation}
that $|\,\psi\,\rangle$ satisfies the differential equation
\begin{equation}
\label{dyn} i\hbar\,\partial_t\,|\,\psi\,\rangle =
\hat{H}\,|\,\psi\,\rangle + i\hbar\,\partial_t\,
\langle\,3\,|\,\psi\,\rangle |\,3\,\rangle \,.
\end{equation}
Consequently, the vector (\ref{dark}) describes correctly the
dynamics of the atom in the lower-level subspace. Therefore, the
atom remains in the dark state (\ref{dark}), as long as the
atom's evolution never leads to an overpopulation at the top
level $|\,3\,\rangle$. The initial relaxation-dominated regime
has prepared the dark state, but later the atom follows
dynamically without relaxation \cite{Fleisch}.

In responding to the light fields, the evolving atoms constitute a
macroscopic dipole density called the matter polarization $P_A$.
Consider a one-dimensional model for light propagation. The matter
polarization influences the probe light according to the wave
equation
\begin{equation}
\label{waveeq} \left(\partial_t^2-c^2\partial_z^2\right) E =
-\varepsilon_0^{-1}\,\partial_t^2 P_A \,.
\end{equation}
Each atom generates a dipole moment of $\frac{1}{2}\kappa_{13}
{\rm tr}\{\hat{\rho}\,|\,1\,\rangle \langle \,3\,|\,\}$,
oscillating at positive frequencies, that contributes to the
total dipole density. Therefore, a medium with $n_A$ atoms per
volume generates a matter polarization with the
positive-frequency component
\begin{eqnarray}
P_A^{(+)} & = & \frac{n_A}{2}\, \kappa_{13}\,\langle\,3\,
|\,\psi\,\rangle\langle\,\psi\,|\,1\,\rangle
\nonumber\\
& = & \frac{n_A}{2}\, \kappa_{13}\,e^{-i\omega_0 t} N_0^4\,
\frac{2}{\Omega_c^*}\,i\partial_t\,\frac{\Omega_p}{\Omega_c}
\nonumber\\
& = &
n_A\,\frac{\kappa_{13}^2}{\hbar}\,\frac{N_0^4}{|\,\Omega_c|^2}
\nonumber\\
& & \times \left(i\partial_t - \omega_0 -
i\frac{(\partial_t|\,\Omega_c|)}{|\,\Omega_c|} + \dot{\theta}_c
\right)E^{(+)}_p
\end{eqnarray}
where $\theta_c = {\rm arg}\Omega_c$. Assume, for simplicity, that
$\Omega_c$ is real. Otherwise we can easily incorporate the phase
$\theta_c$ of the control field in the phase of the electric
field without affecting the wave equation (\ref{waveeq}), as long
as $\theta_c$ varies slowly compared with the optical frequency
$\omega_0$. We define
\begin{equation}
\label{alphaformula} \alpha = n_A
\frac{\kappa_{13}^2}{2\varepsilon_0\hbar}\,
\frac{\omega_0}{|\,\Omega_c|^2} = \frac{n_A}{2}\,
\frac{\kappa_{13}^2}{\kappa_{12}^2}\,
\frac{\hbar\omega_0}{\varepsilon_0\, |E_c|^2} \,,
\end{equation}
which, as we will see shortly, is the group index (\ref{control}).
We get
\begin{equation}
\varepsilon_0^{-1}\,\partial_t^2 P^{(+)} \approx
-N_0^4\alpha\,2\omega_0\left(i\partial_t - \omega_0 -
i\frac{\dot{\alpha}}{2\alpha}\right)E^{(+)}_p \,.
\end{equation}
and approximate
\begin{eqnarray}
2\omega_0(i\partial_t - \omega_0)\,E^{(+)}_p & \approx &
(i\partial_t + \omega_0)(i\partial_t - \omega_0)\,E^{(+)}_p
\nonumber\\
& = & -(\partial_t^2+\omega_0^2)\,E^{(+)}_p \,.
\end{eqnarray}
In this way we obtain from the general wave equation
(\ref{waveeq}) an equation that is valid for both the positive
and the negative frequency component of the probe light,
\begin{equation}
\label{saturation} \left[\partial_t^2 - c^2 \partial_z^2 +
N_0^4\left(\partial_t \alpha\, \partial_t +
\alpha\,\omega_0^2\right)\right] E_p = 0 \,.
\end{equation}
The dark-state dynamics may lead to a non-linear saturation of
the medium, described by the $N_0^4$ factor in the wave equation
(\ref{saturation}). The non-linearity is relevant when the Rabi
frequencies $|\,\Omega_p|$ and $|\,\Omega_c|$ are comparable. When
the probe is significantly weaker than the control light, the
medium responds linearly,
\begin{equation}
\label{ewave} \left(\partial_t(1+\alpha)\partial_t - c^2
\partial_z^2 + \alpha\,\omega_0^2\right) E_p = 0 \,.
\end{equation}
We have derived the wave equation (\ref{wave}). The group index
(\ref{alphaformula}) determines the group velocity $v_g = c/(1 +
\alpha)$. Remarkably, the lower the intensity of the control beam
is, the slower the probe light becomes. Taken to the extreme,
light freezes when the control light is switched off --- a
paradoxical behavior that is only possible in a dynamical regime
\cite{Fleisch}: a control beam of moderate intensity first
captures the probe light, slowing it down, and then, by ramping
down the control intensity, freezes the probe pulse. Equally
paradoxically, the non-linearity of the EIT medium is stronger
the weaker the control beam is. We show in Appendix B that the
unusual non-linear optics in an EIT medium matters in a
slow-light catastrophe.

\section*{Appendix B: Estimations}

In this appendix we estimate the effect of a slow-light
catastrophe, using the experimental parameters of Refs.\
\cite{Liu,Dutton}. We calculate the photon flux in the far field
and estimate the intensity near the horizon. In order to detect
experimentally the quantum radiation of the catastrophe, the flux
should be sufficiently strong. Close to the horizon, even slightly
detuned slow light leaves the absorption-less transparency window
of EIT, unless the medium becomes non-linear. The intensity near
the horizon determines whether the non-linearity or the absorption
dominates.

First we generalize our one-dimensional model of the slow-light
catastrophe to the three dimensions of space in Cartesian
coordinates ${\bf x} = (x, y, z)$. The spatial profile of the
group index shall be uniform in $x$ and $y$ and parabolic in $z$
with the scale $a$. The propagation of slow light is governed by
the wave equation
\begin{equation}
\left[\partial_t^2 - c^2 \nabla^2 +
\frac{a^2}{z^2}\left(\partial_t^2+\omega_0^2\right) \right]
\varphi = 0 \,.
\end{equation}
We find the stationary solutions and normalize them according to
the scalar product
\begin{equation}
\left(\varphi_1,\varphi_2\right) = i \int
\Big(\varphi_1^*\partial_t\varphi_2 - \varphi_2
\partial_t \varphi_1^*\Big)(1 + \frac{a^2}{z^2})\, d^3x \,.
\end{equation}
We obtain the set of modes (\ref{set}) with
\begin{eqnarray}
u_R^\pm & = &
\frac{\Theta(z)}{\lambda_0\sqrt{2c}}\,e^{-\pi\mu/2}\, \sqrt{k_0
z}\,J_{\pm i\mu}(k_zz)\,e^{ik_xx + ik_yy - i\omega t}
\,,\nonumber \\
u_L^\pm({\bf x}) & = & u_R^\mp(-{\bf x})\,, \nonumber\\
w_R^\pm & = & \frac{1}{\sqrt{1-e^{-2\pi\mu}}} \,\left( u_R^\pm -
e^{-\pi\mu}\,u_R^\mp\right)\,,\nonumber\\
w_L^\pm({\bf x}) & = & w_R^\mp(-{\bf x})\,, \label{w3d}
\end{eqnarray}
and $\lambda_0 = 2\pi/k_0$ and $k_0 = \omega_0/c$. Armed with the
three-dimensional modes, we turn to calculating the energy flux.
The Poynting vector of light is the time-averaged expectation
value of the normally ordered Poynting operator \cite{MandelWolf},
\begin{eqnarray}
{\bf P} & = & \lim_{T\rightarrow\infty}\, \frac{1}{2T}
\int_{-T}^{+T} \langle : -\hbar c^2 (\partial_t
\hat{\varphi})(\nabla \hat{\varphi}) : \rangle\,dt \nonumber \\
& = & -\hbar c^2 \int\int \langle : \dot{w}_1\nabla
w_2^*\,\hat{a}_1\hat{a}_2^\dagger + \dot{w}_1^*\nabla
w_2\,\hat{a}_1^\dagger\hat{a}_2 :\rangle \,d^3q_1\, d^3q_2
\nonumber \\
& = & -\hbar c^2 \int \left(\dot{w}\nabla w^* + \dot{w}^*\nabla
w\right) \bar{n}\,d^3q \nonumber\\
& = & \hbar c \int 2 \omega^2 {\bf q} \, |w|^2 \bar{n}\,d^3q \,.
\label{farflux}
\end{eqnarray}
In the last step we have utilized that the $w$ modes approach
plane waves (\ref{hankel}) far away from the horizon. In view of
the narrow bandwidth of EIT we can replace $\omega$ by the
resonance frequency $\omega_0$. Consider a radiating surface with
area $A$ observed from the distance $r$ under the angle
$\vartheta$. The range of wave vectors contributing to the flux
(\ref{farflux}) is restricted to lie within the solid angle of
the surface, $A \cos\vartheta/r^2$.  In the line of sight we thus
get the Poynting-vector component
\begin{eqnarray}
P & = & \hbar\omega_0\,\frac{A\cos\vartheta}{r^2}\,2 c\, \omega_0
\int
|w|^2 \bar{n}\,dq \nonumber\\
& = & \hbar\omega_0\,\frac{A\cos\vartheta}{r^2} \, \omega_0\,
\nonumber\\
& & \times
 \int \frac{1 - e^{-2\pi\mu}}{2\pi}\, \left(e^{\pi\mu}
+ e^{-\pi\mu}\right)^{-2} dq \,,
\end{eqnarray}
having applied the asymptotics (\ref{hankel}). We employ $\mu$ as
the integration variable, with $\mu(\partial\mu/\partial q) = a^2
k_0^2$, and obtain in terms of the critical detuning
(\ref{parameters}) the photon flux
\begin{eqnarray}
\frac{P}{\hbar\omega_0} & = &
\frac{A\cos\vartheta}{r^2\lambda_0^2} \, \omega_0\,\delta_0\,
\nonumber\\
& & \times \frac{4}{\pi} \int_0^\infty\left(1 -
e^{-2\pi\mu}\right) \left(e^{\pi\mu} + e^{-\pi\mu}\right)^{-2}
\mu\, d\mu
\nonumber\\
& = & \frac{A\cos\vartheta}{r^2\lambda_0^2} \,
\omega_0\,\delta_0\,\eta_0 \,,\nonumber\\
\eta_0  & = & \frac{2}{\pi^3}\left(\ln 2 - \frac{\pi^2}{24}\right)
\,.
\end{eqnarray}
The flux integrated over the two half spheres around the radiating
surface gives the total photon-production rate
\begin{equation}
N = 4\pi \int_0^{\pi/2}
\frac{Pr^2}{\hbar\omega_0}\,\sin\vartheta\, d\vartheta =
\frac{A}{\lambda_0^2}\,\omega_0\,\delta_0\,2\pi\eta_0 \,.
\end{equation}
For light in the optical spectral range, $2\pi\omega_0\,\eta_0$
is about $4\times10^{14}{\rm Hz}$. Assuming a critical detuning
$\delta_0$ of $10^{-10}$ a surface of $10^2\lambda_0$ could
generate millions of photons per second. Usually a photodetector
captures only a small solid angle of the radiation emitted from a
localized source, and a detector is not perfectly efficient in
counting all photons. Yet the radiation of the slow-light
catastrophe seems to be strong enough to be detectable.

Let us estimate the strength of the control field needed to
generate the flux we have calculated. We calibrate the field
strength in terms of a Rabi frequency (\ref{rabi}). According to
Eq.\ (\ref{parameters}), a critical detuning $\delta_0$ of
$10^{-10}$ corresponds to a length scale $a \approx 5\times
10^3\lambda_0$ of the group-index profile (\ref{catastrophe}). In
the experiment \cite{Liu} a group index of $10^7$ is generated by
a control field with Rabi frequency $\Omega_c = 2.57\times 2\pi
{\rm MHz} \approx 16 {\rm MHz}$. The group index is inversely
proportional to $\Omega_c^2$, which leads to a Rabi frequency
$\Omega_c$ of about $5\times 10^4\sqrt{\alpha}\, {\rm MHz}$. For
the profile (\ref{catastrophe}) of our slow-light catastrophe we
find that $\Omega_c$ should grow linearly by $10{\rm MHz}$ per
wave-length distance away from the horizon,
\begin{equation}
\Omega_c = 10\,\frac{z}{\lambda_0}\,{\rm MHz} \,.
\end{equation}
Probably $10$ to $100$ wave lengths are sufficient to establish
the wave catastrophe. This would take a continuous-wave control
field with a maximal Rabi frequency of $10^2$ to $10^3 {\rm MHz}$.

The ratio $\Omega_p/\Omega_c$ of the Rabi frequencies
(\ref{rabi}) determines whether non-linear effects dominate near
the horizon. We calculate $\Omega_p$ using the results of the
linear theory. Strictly speaking, Rabi frequencies refer to
classical fields. Here we regard the time-averaged and normally
ordered expectation value of the quantum intensity $\hat{E}_p^2$
as being proportional to $\Omega_p^2$. Comparing Eq.\ (1) of Ref.
\cite{Dutton} with our Eq.\ (\ref{alphaformula}) we find a
relation between the Rabi frequency and the intensity,
\begin{equation}
\Omega_p^2 = 3\times10^3
\langle\frac{\varepsilon_0\hat{E}_p^2}{\hbar\omega_0}\rangle
\,{\rm m}^3{\rm Hz}^2 \,,
\end{equation}
for the atomic transition employed in the experiments
\cite{Liu,Dutton}. We express $\hat{E}_p$ in terms of the field
$\hat{\varphi}$ in units of the vacuum noise, and get
\begin{eqnarray}
\langle\frac{\varepsilon_0\hat{E}_p^2}{\hbar\omega_0}\rangle & = &
\lim_{T\rightarrow\infty}\, \frac{1}{2T} \int_{-T}^{+T} \omega_0\,
\langle : \hat{\varphi}^2 : \rangle\,dt \nonumber\\
& = & 2\omega_0 \int |w|^2\,\bar{n}\,d^3q \,.
\end{eqnarray}
Close to the horizon the $w$ waves obey the asymptotics
\begin{equation}
|w|^2 \sim
\frac{k_0z}{2c\lambda_0^2}\,\frac{1-e^{-\pi\mu}}{1+e^{-\pi\mu}}\,
e^{-\pi\mu}\,\frac{{\rm sinh}(\pi\mu)}{\pi\mu} \,,
\end{equation}
as we find from the definition (\ref{w3d}) and the behavior
(\ref{power}) of the Bessel functions, utilizing Eq.\ 1.2.(6) of
Ref.\ \cite{Erdelyi1} for the gamma function
$x!\equiv\Gamma(x+1)$. We obtain
\begin{eqnarray}
\langle\frac{\varepsilon_0\hat{E}_p^2}{\hbar\omega_0}\rangle & =
& 2\omega_0\,2\pi\int_0^\infty |w|^2
\bar{n}\,\frac{\mu}{a^2k_0^2}\,d\mu \nonumber\\
& = & 32\pi(1-\ln 2)\,\delta_0\, \frac{z}{\lambda_0^4}
\end{eqnarray}
and, consequently,
\begin{equation}
\Omega_p \sim 8.6\,\sqrt{\frac{z}{\lambda_0}}\,{\rm MHz} \,.
\end{equation}
Therefore, the Rabi frequencies $\Omega_p$ and $\Omega_c$ are
comparable at about half a wave-length distance away from the
horizon. Here a critical detuning of $10^{-10}$ lies still within
the transparency window generated by a control field of $5{\rm
MHz}$ Rabi frequency. The EIT medium becomes non-linear.

\end{document}